\documentclass[11pt]{llncs}

\usepackage{algorithm2e}
\usepackage{amsfonts}
\usepackage{amsmath}
\usepackage{booktabs}
\usepackage{pgfplots}
\usepackage{tabularx}
\usepackage{tikz}
\usepackage{url}
\usepackage{wrapfig}
\usepackage[strings]{underscore}

\input{kvmacros}

\mathchardef\mhyphen="2D

\begin{document}

\title{Exact Synthesis of ESOP Forms}

\author{%
\!Heinz Riener\inst{1}\!\and\!R\"udiger Ehlers\inst{2}\!\and\!Bruno Schmitt\inst{1}\!\and\!Giovanni\ De\ Micheli\inst{1}
}%
\institute{
EPFL, Lausanne, Switzerland\\
\and
University of Bremen, Bremen, Germany}
\maketitle

\begin{abstract}
We present an exact synthesis approach for computing \emph{Exclu\-sive-or Sum-of-Products}~(ESOP) forms with a minimum number of product terms using Boolean satisfiability.  Our approach finds one or more ESOP forms for a given Boolean function.  The approach can deal with incompletely-specified Boolean functions defined over many Boolean variables and is particularly fast if the Boolean function can be expressed with only a few product terms.  We describe the formalization of the ESOP synthesis problem with a fixed number of terms as a decision problem and present search procedures for determining ESOP forms of minimum size.  We further discuss how the search procedures can be relaxed to find ESOP forms of small sizes in reasonable time.  We experimentally evaluate the performance of the SAT-based synthesis procedures on completely- and incompletely-specified Boolean functions.
\end{abstract}

\section{Introduction}
In the design of \emph{Very Large Scale Integration}~(VLSI) systems, two-level logic representations are classically used to represent and manipulate Boolean functions.  \emph{Exclusive-or Sum-of-Products}~(ESOP) is a two-level normal form representation of a Boolean function that consists of one level of multi-input AND gates followed on the next level by one multi-input XOR gate.  ESOP forms play an important role in logic synthesis due to their improved compactness for arithmetic or communication circuits with respect to other two-level representations~\cite{Sasao96} and their excellent testability properties~\cite{KalayHP00}.  The inherent reversibility of the XOR operation, moreover, makes ESOP forms particularly suitable in applications such as security~\cite{MizukiOS07,KolesnikovS08} or quantum computation~\cite{FazelTR07}.

The ESOP representation of a Boolean function is not unique, i.e., the same Boolean function can be expressed as multiple structurally different, but semantically equivalent ESOP forms.  In practice, it is important to find a small representation of an ESOP form to reduce the overall costs for realizing it in hardware or implementing it in software.  The problem of synthesizing an ESOP form for a given Boolean function is to identify a set of product terms over the Boolean variables of the function such that each minterm in the OFF-set of the function is covered by the product terms an even number of times and each minterm in the ON-set of the Boolean function is covered an odd number of times.

Finding ESOP forms with a small or a minimum number of product terms is hard and numerous exact and heuristic synthesis methods~\cite{MishchenkoP01,Gaidukov02,StergiouP04,SampsonKVP12,Papakonstantinou14,Papakonstantinou18} for solving this problem have been proposed.  Heuristic methods focus on finding small (but not necessarily minimum) ESOP forms; they are fast, but only examine a subset of the possible search space.  Heuristic methods, e.g., the Exorcism approach~\cite{MishchenkoP01}, usually operate in two phases.  In the first phase, an ESOP form with a sub-optimal number of product terms is derived from the Boolean function, e.g., by translating each minterm of the Boolean function into one product term or translating the function into special cases of ESOP forms such as pseudo-Kronecker expressions~\cite{Drechsler99}.  In the second phase,  the ESOP form is iteratively optimized and reshaped using cube transformations with the overall goal of merging as many product terms as possible.  The cube transformations are applied to each pair of product terms that potentially lead to merging them or with other product terms of the ESOP form.  The second phase terminates when, after several iterations, no further size reduction is achieved.  Heuristic methods produce small ESOP forms in reasonable time, but suffer from local minima that cannot be easily escaped.  In contrast, exact methods find an ``exact'' ESOP form, i.e., an ESOP form with a minimum number of product terms, but either require to store large tables of pre-computed information~\cite{SampsonKVP12,Papakonstantinou14} or suffer from tremendously high runtimes~\cite{Papakonstantinou18}.  For instance, the tabular-based methods described by Gaidukov~\cite{Gaidukov02} or Papakonstantinou~\cite{Papakonstantinou14} require pre-computed tables of all exact ESOP forms for Boolean functions over $n-1$ Boolean variables to derive an exact ESOP form for a Boolean function over $n$ Boolean variables.  Due to the exponential growth of the number of Boolean functions with the number of Boolean variables, these methods become too time and memory consuming when $n > 6$.  Alternative exact synthesis approaches such as a recent formulation of the ESOP synthesis problem using non-linear programming~\cite{Papakonstantinou18}, can take hundreds of hours for synthesizing a single exact ESOP form.

Until today, a large gap between the number of product terms optimized with heuristic methods and exact methods remains.  Where exact methods hardly can deal with more than $8$ Boolean variables and a few product terms, heuristic methods nowadays, e.g., in the quantum domain, have to deal with the optimization of ESOP forms with $10^5$ or $10^6$ products terms over $16$ and more Boolean variables~\cite{SoekenRW+17}.  Our experiments with large-scale ESOP forms showed that heuristic optimization method can often achieve a reduction of $50-80\%$ in the number of ESOP terms with respect to the size of the initial ESOP form.  Due to the large combinational search space of the ESOP synthesis problem, lower bounds on the number of required product terms are only known for Boolean functions with a few Boolean variables, such that the capabilities of ESOP optimization techniques remain unclear.

In this paper, we investigate the exact synthesis of ESOP forms using Boolean satisfiability (SAT).  SAT-based approaches are very successful on a variety of different verification and synthesis problems.  We present an exact synthesis approach for computing ESOP forms with a minimum number of product terms.  Starting from a specification in form of a possibly incompletely-specified Boolean function, our approach iteratively constructs a Boolean constraint satisfaction problem that is satisfiable if and only if an ESOP form with $k$ (initially $k = 1$) product terms that implements the specification exists.  The problem is then solved utilizing a SAT-solver and, if satisfiable, an ESOP form with $k$ product terms is returned.  Otherwise, if unsatisfiable, $k$ is increased and the synthesis process is restarted.  The synthesis approach is hardly affected by the number of Boolean variables and particularly fast if the Boolean function can be expressed by using only a few product terms.  We argue that such a SAT-based exact synthesis procedure can be a backbone of a new generation of heuristic ESOP optimization methods that, instead of relying on cube transformations applied to a pair of product terms, are capable of optimizing small subsets (windows) of product terms.

The proposed approach is the first ESOP synthesis technique based on Boolean satisfiability.  We further present a relaxation of the technique to compute ESOP forms with size close to minimum leveraging the SAT-solver's conflict limit.  We have implemented SAT-based exact synthesis for ESOPs and the relaxation of the approach using an off-the-shelf SAT-solver and show in the experiments that SAT-based ESOP synthesis can be readily used to synthesize ESOP forms with up to $8$ Boolean variables and up to $100$ terms.  As benchmarks, we use completely-specified Boolean functions that are used as representatives of the NPN$4$ equivalence classes~\cite{GotoT62} as well as completely-specified Boolean functions that appeared in technology mapping using look-up tables (LUTs) with at most $8$ inputs ($8$-LUT mapping).  Moreover, we use a set of randomly-generated incompletely-specified Boolean functions with up to $8$ Boolean variables.

\section{Background}\label{sec:background}

\textbf{Exclusive-or Sum-of-Products (ESOP).}
Let $\mathbb{B} = \{0,1\}$ and $\mathbb{B}_{3} = \{0,1,-\}$ with the third element `$-$' which denotes don't care.  An \emph{ESOP form} in $n$ Boolean variables $x_1, \ldots, x_n$ is a Boolean expression
\begin{align}\label{eq:ESOP}
  \bigoplus_{j=1}^k \left( \bigwedge_{i=1}^n x_i^{l_{i,j}} \right),
\end{align}
where the operators $\oplus$ and $\land$ denote standard addition (XOR) and multiplication (AND) in the Galois field with two-elements, respectively, each $l_{i,j} \in \mathbb{B}_{3}$ is a constant and each expression
\begin{align}
  x_i^{l_{i,j}} = \begin{cases}\bar{x}_i,&\text{if }l_{i,j} = 0\\x_i,&\text{if }l_{i,j} = 1\\1,&\text{if }l_{i,j} = -\end{cases}
\end{align}
for $1 \leq i \leq n$ and $1 \leq j \leq k$.  We say that $k$ is the \emph{size} of the ESOP form and call each conjunction $x_1^{l_{1,j}} \cdots x_n^{l_{n,j}}$, $1 \leq j \leq k$, that appears in the ESOP form a \emph{product term}.  The Boolean expression in (\ref{eq:ESOP}) is often compactly notated as a list of words
\begin{align}
  l_{1,1} \cdots l_{n,1} \quad l_{1,2} \cdots l_{n,2}  \qquad \dots \qquad l_{1,k} \cdots l_{n,k},
\end{align}
where each word $l_{1,j} \cdots l_{n,j}$ is of fixed length $n$.

\noindent\textbf{Distance of product terms.}
Suppose that
\begin{align}
  u = x_1^{l_{1,p}} \cdots x_n^{l_{n,p}}\qquad{}\text{ and }\qquad{}v = x_1^{l_{1,q}} \cdots x_n^{l_{n,q}}.
\end{align}
are two product terms in $n$ Boolean variables.  We define the \emph{distance} $d(u,v)$ of $u$ and $v$ as the number of different $l_{i,j}$ for $1 \leq i \leq n$ and $j \in \{p,q\}$, i.e.,
\begin{align}
  d(u,v) = \sum_{i=1}^n [ l_{i,p} \neq l_{i,q} ],
\end{align}
where $[.]$ denote the Iverson brackets.  We say if $d(u,v) = m$, that $u$ and $v$ have distance $m$ or are $m$-distant.

\noindent\textbf{ESOPs describing Boolean functions.}
An ESOP form semantically describes a (single-output) Boolean function $f : \mathbb{B}^n \rightarrow \mathbb{B}$, which maps assignments of the Boolean variables $x_1, \dots, x_n \in \mathbb{B}$ to truth values $f(x_1, \dots, x_n) \in \mathbb{B}$.  Each assignment to all Boolean variables $x_1, \dots, x_n$ is called a \emph{minterm} and can be interpreted as the decimal number $\sum_{i=1}^n x_i 2^{i-1}$ when read as $(x_n \cdots x_1)_2$.

A completely-specified Boolean function $f : \mathbb{B}^n \rightarrow \mathbb{B}$ in $n$ Boolean variables can be uniquely represented as a truth table, i.e., a word $b_{2^n} \cdots b_1$ of length $2^n$, where $b_j = f(j-1)$ for $1 \leq j \leq 2^n$.  An incompletely-specified Boolean function $g : \mathbb{B}^n \rightarrow \mathbb{B}_3$ can be represented by two completely-specified Boolean functions $f : \mathbb{B}^n \rightarrow \mathbb{B}$ and $c : \mathbb{B}^n \rightarrow \mathbb{B}$, where $f(x) = [g(x) = 1]$ and $c(x) = [g(x) \neq -]$.  We call $c$ the care function of $g$.

Two ESOP forms are semantically \emph{equivalent} if they describe the same Boolean function.  An ESOP form with size $k$ is \emph{minimum} if and only if no semantically equivalent ESOP form with less product terms exists.  Minimum ESOP forms are in general not unique.

\section{SAT-based Exact ESOP Synthesis}\label{sec:synthesis}

\subsection{Exact synthesis of ESOP forms}

\textbf{Objective.} We aim for synthesizing minimum ESOP forms in $n$ Boolean variables when a completely-specified Boolean function or incompletely-specified Boolean function is provided as specification.  In case of com\-plete\-ly-specified Boolean functions, this objective can be formally described as follows: given a single-output Boolean function $f : \mathbb{B}^n \rightarrow \mathbb{B}$ over $n$ Boolean variables $x_1, \dots, x_n$, find an integer $k$ and constants $l_{i,j} \in \mathbb{B}_3$ for $1 \leq i \leq n$ and $1 \leq j \leq k$ such that
\begin{align}\label{eq:objective}
  \bigoplus_{j=1}^k \left( \bigwedge_{i=1}^n x_i^{l_{i,j}} \right) = f(x_1,\dots,x_n)\text{ for all }x_1, \dots, x_n
\end{align}
and $k$ is minimum.  The case of incompletely-specified Boolean functions can be addressed similarly to (\ref{eq:objective}).

\begin{example}\label{ex:intro}
  As an introductory example, consider the in\-com\-ple\-tely-spe\-ci\-fied Boolean function described by the truth table \texttt{0x688C802028222222}\footnote{We use hexadecimal notation to shorten the string representation of the (binary) truth tables of Boolean functions.} over $6$ Boolean variables with care function \texttt{0x6AAEFF3FFEBFEAA6}.  A minimum ESOP form, for instance, is
  \begin{align}
    \bar{x}_1 x_3 \bar{x}_4 \bar{x}_5 x_6 \oplus
    \bar{x}_1 x_2 \bar{x}_3 x_5 \bar{x}_6 \oplus
    \bar{x}_1 \bar{x}_3 \bar{x}_4 \bar{x}_6 \oplus
    \bar{x}_2 \bar{x}_5 \bar{x}_6 \oplus
    \bar{x}_1 x_2 x_6,
  \end{align}
  which requires $5$ product terms and can be equivalently written as
  \begin{align}
    {\tt 0\mhyphen{}1001} \quad {\tt 0\mhyphen{}00\mhyphen{}0} \quad {\tt \mhyphen{}0\mhyphen{}\mhyphen{}00} \quad {\tt 010\mhyphen{}10} \quad {\tt 01\mhyphen{}\mhyphen{}\mhyphen{}1}.
  \end{align}

  In general, minimum ESOPs are not unique.  The same Boolean function may also be represented as the ESOP form
  \begin{align}
    {\tt 0\mhyphen{}1001} \quad {\tt 0100\mhyphen{}0} \quad {\tt \mhyphen{}0\mhyphen{}\mhyphen{}00} \quad {\tt 0\mhyphen{}0\mhyphen{}10} \quad {\tt 01\mhyphen{}\mhyphen{}\mhyphen{}1}
  \end{align}
  or
  \begin{align}
    {\tt 0\mhyphen{}1001} \quad {\tt 0\mhyphen{}00\mhyphen{}0} \quad {\tt \mhyphen{}\mhyphen{}\mhyphen{}\mhyphen{}00} \quad {\tt 011\mhyphen{}10} \quad {\tt 01\mhyphen{}\mhyphen{}\mhyphen{}\mhyphen{}}.
  \end{align}
\end{example}

Finding minimum ESOP forms is, due to the large combinational search space, a challenging problem.  In~\cite{Papakonstantinou18}, a minimum ESOP form for the Boolean function in the previous example was found in roughly $18$h using integer non-linear programming and Matlab as a solving engine.  The authors, moreover, point out that decomposition-based ESOP synthesis approaches,\ e.g.,~\cite{SampsonKVP12}, require up to $4$h for synthesizing minimum ESOP forms for incompletely-specified Boolean functions over $6$ Boolean variables.

\subsection{SAT-based exact synthesis procedure}

In this section, we propose a SAT-based exact synthesis approach for ESOP forms.  The approach is based on ideas from Kamath et al.~\cite{KamathKRR92} and our previous work on learning two-level patches to correct combinational Boolean circuits~\cite{RienerEF17}.  Our approach synthesizes an ESOP form for the Boolean function in Example~\ref{ex:intro} in less than a second.  We formalize the search problem as a series of Boolean constraint satisfaction problems---one for each possible ESOP size $k$ (starting with $k = 1$) and employ a decision procedure for Boolean satisfiability to decide the satisfiability of the constraints.  The constraints are constructed in such a way that they are satisfiable if and only if an ESOP form with $k$ product terms exists and each satisfying assignment corresponds to an ESOP form with $k$ product terms.  If the constraints are unsatisfiable, then no ESOP form restricted to $k$ product terms, that is equivalent to the provided Boolean function, exists.  By systematically solving the constraint satisfaction problem for increasing values of the size parameter $k$, a minimum ESOP form is guaranteed to be found.

\noindent\textbf{Formulation of the constraint satisfaction problem.}
Suppose that $f : \mathbb{B}_3^n \rightarrow \mathbb{B}$ is a (single-output) Boolean function over $n$ Boolean variables.  We formulate the problem of finding an ESOP form equivalent to $f$ with $k$ product terms as a constraint satisfaction problem in propositional logic using $2nk$ Boolean variables, $p = p_{1,1}$, \dots, $p_{k,n}$ and $q = q_{1,1}, \dots, q_{k,n}$, where $n$ is the number of Boolean variables of $f$, $k$ is the size of the ESOP form, and
\begin{align}\label{eq:variables}
  p_{j,l} = [\text{$x_l$ in product term $j$}] \quad \text{ and } \quad q_{j,l} = [\text{$\bar{x}_l$ in product term $j$}]
\end{align}
for $1 \leq j \leq k$ and $1 \leq l \leq n$.

For each assignment $x_1 \cdots x_n \in \mathbb{B}_3^n$ of the Boolean function $f$ with the corresponding output value $f(x_1,\ldots,x_n) = b$, we introduce $k$ auxiliary Boolean variables $z = z_1,\dots,z_k$ and add $k \cdot n + k$ clauses
\begin{align}\label{eq:clauses}
  \bigwedge_{j=1}^k \bigwedge_{l=1}^{n} \bigl( \bar{z}_j \vee \mathrm{ITE}(x_i,\bar{q}_{j,l}, \bar{p}_{j,l}) \bigr)\quad\text{ and }\quad
  \bigwedge_{j=1}^k \left( z_j \vee \bigvee_{l=1}^{n} \mathrm{ITE}(x_i, q_{j,l}, p_{j,l}) \right),
\end{align}
which ensure that if $z_j = 1$ then the $j$-th product term evaluates to $1$ for assignment $x_1 \cdots x_n$ and if $z_j = 0$ then the $j$-th product term evaluates to $0$ for assignment $x_1 \cdots x_n$.  The \emph{if-then-else}-operator is defined as
\begin{align}
  \mathrm{ITE}(x_i, v_{j,l}, u_{j,l}) = \begin{cases}v_{j,l}, &\text{ if }x_i = 1\\u_{j,l}, &\text{ if }x_i = 0\\false, &\text{ otherwise}\end{cases}
\end{align}
with $v_{j,l} \in \{q_{j,l}, \bar{q}_{j,l}\}$ and $u_{j,l} \in \{p_{j,l}, \bar{p}_{j,l}\}$, respectively.  One additional XOR-constraint
\begin{align}\label{eq:xor constraints}
  \left( \bigoplus_{j=1}^k z_j \right) = b
\end{align}
per assignment guarantees that an odd number of $z_j$s evaluates to $1$ if $b = 1$ and an even number if $b = 0$.

This constraint satisfaction problem is satisfiable if and only if an ESOP form of size~$k$ exists and each satisfying assignment $\hat p_{1,1}, \dots, \hat p_{k,n}$ and $\hat q_{1,1}, \dots, \hat q_{k,n}$ corresponds to one possible implementation.

\noindent\textbf{Translating XOR-constraints to CNF.}  All XOR-constraints in the constraint satisfaction problem are, by construction, formulated over disjoint sets of Boolean variables such that techniques like Gaussian elimination are not effective.  Instead, we translate each XOR-constraint first into an equivalent XOR-clause by flipping one of the Boolean variables if and only if $b = 0$, i.e.,
\begin{align}
  (z_1 \oplus \dots \oplus z_k) = b \quad\Longrightarrow\quad \begin{cases}z_1 \oplus \dots \oplus z_k,&\text{if }b = 1\\z_1 \oplus \dots \oplus \bar{z}_k,&\text{if }b = 0.\end{cases}
\end{align}
Then, we select two literals $l_a, l_b$ from the XOR-clause and apply the Tseitin transformation to generate four clauses $(\bar{z}_a \vee \bar{z}_b \vee \bar{u})$, $(z_a \vee z_b \vee \bar{u})$, $(z_a \vee \bar{z}_b \vee u)$, $(\bar{z}_a \vee z_b \vee u)$ with the newly introduced Boolean variable $u$ and repeat this process until only one literal is left which is added as a unit clause.

\begin{algorithm}[t]
  \SetKwInOut{Input}{input}\SetKwInOut{Output}{output}
  \SetKwFunction{MakeCSP}{MakeCSP}\SetKwFunction{MakeESOP}{MakeESOP}\SetKwFunction{SAT}{SAT}
  \Input{a (possibly incompletely-specified) Boolean function~$f$}
  \Output{a minimum ESOP functionally equivalent to $f$}
  \BlankLine
  \For{$k \leftarrow 1, 2, \dots$}{%
    $\varphi(p,q,z)$ $\leftarrow$ \MakeCSP{$k$,$f$}\;
    \If{$\hat{p}, \hat{q}$ $\models$ \SAT{$\exists z : \varphi(p,q,z)$}}{%
      \Return \MakeESOP{$\hat{p}$,$\hat{q}$}\;
    }%
  }%
  \caption{SAT-based exact ESOP synthesis}
  \label{alg:exact esop}
\end{algorithm}

\noindent\textbf{SAT-based exact ESOP synthesis.}
The overall exact synthesis procedure is sketched in Algorithm~\ref{alg:exact esop}.  The function $\texttt{MakeCSP}$ constructs the constraint satisfaction problem $\varphi$ in the Boolean variables $p, q, z$ for a given Boolean function $f$ and size parameter $k$ as described above.  The function $\texttt{SAT}$ refers to the invocation of a decision procedure for the Boolean satisfiability problem, usually called a \emph{SAT-solver}, and is assumed to decide the satisfiability of $\varphi$ and, if satisfiable, to also provide a satisfying assignment $\hat p$ and $\hat q$ for variables $p$ and $q$.  The assignment to the intermediate Boolean variables $z$ is for the construction of no further interest and not returned.  Finally, the function $\texttt{MakeESOP}$ constructs an ESOP form from the assignment $\hat p$ and $\hat q$ according to the rules described in (\ref{eq:variables}).  Note that~Algorithm~\ref{alg:exact esop} always terminates, but may run out of resources (memory or time) if the minimum ESOP requires many product terms.  Thus in practice usually an additional termination criterion in form of an upper bound for the size parameter $k$ or maximum number of conflicts examined by the SAT-solver is provided.

\begin{algorithm}[t]
  \SetKwInOut{Input}{input}\SetKwInOut{Output}{output}
  \SetKwFunction{MakeESOP}{MakeESOP}\SetKwFunction{SAT}{SAT}\SetKwFunction{AddConstraints}{AddConstraints}\SetKwFunction{NotEquivalent}{NotEquivalent}
  \Input{a (possibly incompletely-specified) Boolean function~$f$}
  \Output{a minimum ESOP $r$ functionally equivalent to $f$}
  \BlankLine
  $r \leftarrow \epsilon$\;
  $k \leftarrow 1$\;
  $\varphi(p,q,z) \leftarrow \textsf{true}$\;
  \While{$m$ $\leftarrow$ \NotEquivalent{$f$,$r$}}{%
    $\varphi \leftarrow$ \AddConstraints{$\varphi$,$m$}\;
    \eIf{$\hat{p}, \hat{q}$ $\models$ \SAT{$\exists z : \varphi(p,q,z)$}}{%
      r $\leftarrow$ \MakeESOP{$\hat{p}$,$\hat{q}$}\;
    }{%
      $r \leftarrow \epsilon$\;
      $k \leftarrow k + 1$\;
      $\varphi(p,q,z) \leftarrow \textsf{true}$\;
    }%
  }%
  \Return{$r$}\;
  \caption{SAT-based exact synthesis guided by counterexamples}
  \label{alg:exact cegar esop}
\end{algorithm}

\noindent\textbf{Counterexample-guided abstraction-refinement.}
Algorithm~\ref{alg:exact esop} synthesizes an ESOP form in one step.  Alternatively, counterexample-guided abstraction-refinement can be employed as shown in Algorithm~\ref{alg:exact cegar esop}.  The idea of the abstraction-refinement loop is to iteratively update a candidate ESOP form $r$ (starting from the empty ESOP form $\epsilon$) until it eventually becomes semantically equivalent to the Boolean function $f$ to be synthesized.  In each iteration, the constraints of one assignment $x = x_1 \cdots x_n$ for which $r$ and $f$ evaluate differently ($r(x) \neq f(x)$) are added (\texttt{AddConstraints}) to the constraint satisfaction problem and $r$ is resynthesized.  If $\varphi$ becomes unsatisfiable, then the constraints cannot be solved within the current restriction to $k$ product terms and $k$ needs to be relaxed.  If $f$ and $r$ are equivalent, i.e., no counterexample $x = x_1 \cdots x_n$ is found by \texttt{NotEquivalent}, then $r$ is returned as an ESOP form semantically equivalent to $f$.  The main advantage of Algorithm~\ref{alg:exact cegar esop} over Algorithm~\ref{alg:exact esop} lies in its ability to abstract from unnecessary constraints which keeps the constraint satisfaction problem as small as possible.  The algorithm is fast mainly because modern backtrack search-based SAT-solvers support \emph{incremental solving}~\cite{EenS03} and are able to maintain learned information when new constraints are added to a satisfiability problem.  The oracle \texttt{NotEquivalent} has to be capable of verifying whether a candidate ESOP form $r$ is functionally equivalent to the Boolean function $f$.  For Boolean functions with up to $16$ Boolean variables, simulation using explicit representations such as truth tables can be done very quickly.  For Boolean functions with more than $16$ Boolean variables, BDD- or SAT-based procedure can be employed.

\subsection{Extensions and variations}

\noindent\textbf{Downward vs. upward search.}
Algorithm~\ref{alg:exact cegar esop} describes an upward search procedure to find a minimum ESOP form starting with $1$ term.  This approach can be easily modified into a downward search by starting from a maximum number of terms $\hat k$ and iteratively decreasing the number of terms by $1$ as long as the constraint system is satisfiable.  If the constraint system becomes unsatisfiable for a certain number $k$ of terms, the previous $k+1$ terms correspond to a minimum ESOP form.  In practice downward and upward search procedures are useful.  An upward search procedure is fast if the expected minimum $k$ is small.  Otherwise, proving unsatisfiability with a SAT-solver becomes too time consuming.  A downward search procedure is fast if the expected minimum $k$ is close to the initially provided term limit $\hat k$.

\noindent\textbf{Conflict limit.}
For a SAT-solver proving unsatisfiability of a set of constraints, i.e., showing that no assignment exists that satisfies the constraints, often requires labor-intensive analysis.  If the search space is sufficiently large, these proofs are often not completed within reasonable time.  Most modern SAT-solver provide a conflict limit to allow a user to specify a maximum number of possible solving attempts.  If the SAT-solver is unable to find a satisfying assignment within the given conflict limit, the solver reports `unknown' as solution.  In this case, the synthesis algorithm can choose to increase or decrease the current $k$ hoping that the next $k$ is easier to solve because the corresponding constraint system is less or more constrained, respectively.  When a conflict limit is employed in Algorithm~\ref{alg:exact cegar esop}, due to the possible `unknown` solutions, a minimum ESOP form may not be found.  However, in case of a downward search, that systematically decreases $k$, an intermediate `unknown` solution for $k_1$ can be safely ignored if the constraint system is later proved satisfiable for $k_2 < k_1$; whereas in case of an upward search, an intermediate `unknown` solution for $k_1$ can be ignored if the constraint system is proved unsatisfiable for a later $k_2 > k_1$.

\section{Experimental Evaluation}

We have implemented Algorithm~\ref{alg:exact cegar esop} in \emph{easy}, an open-source toolkit for manipulating ESOP forms\footnote{easy, \url{https://github.com/hriener/easy}} using the prominent state-of-the-art SAT-solver Glucose~4.1~\cite{AudemardS18} as decision procedure for Boolean satisfiability. %and BreakId 2.3~\cite{Devriendt0BD16} for static symmetry breaking.

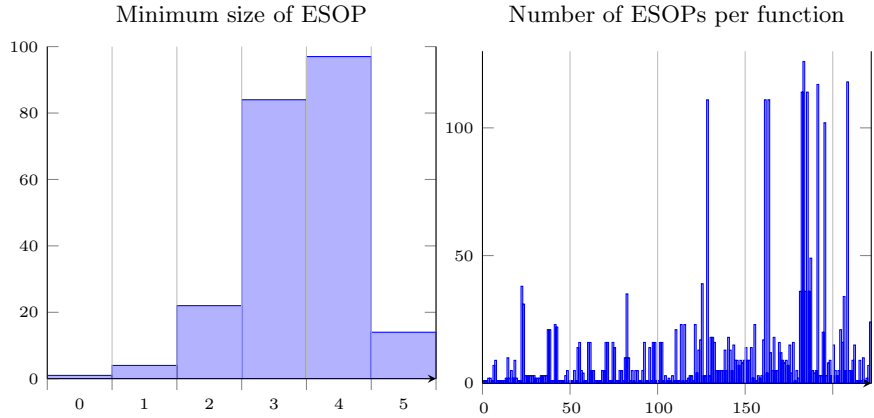
\begin{figure}[t]
  \begin{center}
    \pgfplotsset{every tick label/.append style={font=\tiny}}

    \begin{tikzpicture}
      \begin{axis}[name=Histogram1, title={Minimum size of ESOP}, axis x line=bottom, xmin=0, xmax=6, ybar interval, ymin=0, ymax=100, area style, width=6.75cm, height=6cm]
        \addplot coordinates { (0,1) (1,4) (2,22) (3,84) (4,97) (5,14) (6,0) }; % sentinel (6,0)
      \end{axis}

      \begin{axis}[name=Histogram2, title={Number of ESOPs per function}, at={(Histogram1.outer north east)}, anchor=outer north west, axis x line=bottom, xmin=0, xmax=222, ybar interval, ymin=0, ymax=130, area style, xtick={}, ytick={}, width=6.75cm, height=6cm, x tick label style={xshift=-16pt}]
        \addplot coordinates { (0,1) (1,1) (2,1) (3,2) (4,2) (5,1) (6,7) (7,9) (8,1) (9,1) (10,1) (11,1) (12,1) (13,2) (14,10) (15,2) (16,5) (17,2) (18,9) (19,2) (20,1) (21,1) (22,38) (23,31) (24,3) (25,3) (26,3) (27,3) (28,3) (29,3) (30,2) (31,2) (32,2) (33,3) (34,3) (35,3) (36,3) (37,21) (38,21) (39,1) (40,1) (41,23) (42,22) (43,1) (44,1) (45,1) (46,1) (47,3) (48,5) (49,1) (50,1) (51,1) (52,5) (53,1) (54,14) (55,16) (56,5) (57,4) (58,1) (59,1) (60,16) (61,16) (62,5) (63,5) (64,1) (65,1) (66,1) (67,1) (68,5) (69,5) (70,16) (71,16) (72,1) (73,1) (74,16) (75,14) (76,1) (77,1) (78,5) (79,5) (80,1) (81,10) (82,35) (83,10) (84,1) (85,5) (86,1) (87,5) (88,5) (89,1) (90,1) (91,5) (92,16) (93,1) (94,1) (95,14) (96,1) (97,16) (98,16) (99,1) (100,1) (101,16) (102,16) (103,1) (104,3) (105,1) (106,2) (107,1) (108,3) (109,1) (110,21) (111,1) (112,1) (113,23) (114,1) (115,23) (116,1) (117,2) (118,3) (119,3) (120,1) (121,23) (122,4) (123,13) (124,17) (125,39) (126,3) (127,3) (128,111) (129,3) (130,18) (131,18) (132,16) (133,5) (134,5) (135,5) (136,5) (137,5) (138,13) (139,5) (140,18) (141,13) (142,5) (143,15) (144,9) (145,9) (146,7) (147,9) (148,8) (149,9) (150,14) (151,9) (152,9) (153,14) (154,2) (155,23) (156,1) (157,3) (158,2) (159,3) (160,17) (161,111) (162,4) (163,111) (164,12) (165,5) (166,18) (167,5) (168,5) (169,16) (170,12) (171,9) (172,8) (173,9) (174,7) (175,15) (176,4) (177,16) (178,1) (179,5) (180,2) (181,36) (182,114) (183,126) (184,36) (185,114) (186,36) (187,49) (188,5) (189,4) (190,5) (191,117) (192,3) (193,3) (194,20) (195,102) (196,3) (197,8) (198,9) (199,3) (200,15) (201,5) (202,3) (203,5) (204,21) (205,16) (206,34) (207,5) (208,118) (209,5) (210,5) (211,9) (212,15) (213,1) (214,1) (215,9) (216,1) (217,10) (218,1) (219,2) (220,7) (221,24) (222,0)}; % sentinel (222,0)
      \end{axis}
    \end{tikzpicture}
  \end{center}
  \caption{Synthesis of minimum ESOP forms for NPN$4$.}
  \label{fig:npn4}
\end{figure}

\begin{figure}
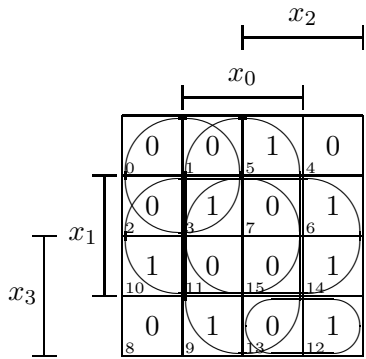

\begin{center}
  % (~x0*~x1*~x2*~x3)⊕(~x2*~x3)⊕(~x1*x2*x3)⊕(x1)⊕(x0)
  \karnaughmap{4}{}%
    {{$x_3$}{$x_2$}{$x_1$}{$x_0$}}%
    {%
      0001 0110
      0110 1010
    }%
    {%
      \put(2,2){\oval(1.9,3.9)[]} % x0
      \put(2,2){\oval(3.9,1.9)[]} % x1
      \put(1,3){\oval(1.9,1.9)[]} % -x2*-x3
      \put(2,2){\oval(1.9,1.9)[]} % -x1*x2*x3
      \put(3,0.5){\oval(1.9,0.9)[]} % -x0*-x1*x2*x3
    }
\end{center}
\caption{Karnaugh map of \texttt{0x166A}.}
\label{fig:kvmap}
\end{figure}

\noindent We have evaluated the SAT-based synthesis approach in three experiments\footnote{The benchmarks and a detailed evaluation of the synthesis results can be found at \url{https://hriener.github.io/misc/2018_easy.html}}:
\begin{enumerate}
\item NPN$4$: We synthesized all ESOP forms of minimum size for the representatives of the NPN$4$ equivalence class.
\item LUT mapping: We synthesized one ESOP form of fixed size and one of minimum size for the Boolean functions that occurred during LUT mapping of Boolean networks.
\item Random: We synthesized one ESOP form of fixed size and one of minimum size for randomly generated Boolean functions.
\end{enumerate}
All experiments have been conducted on a Intel\textsuperscript{\textregistered} Core\textsuperscript{\texttrademark} i7-7567U CPU @ $3.50$GHz with $16$GB RAM.

\noindent\textbf{Correctness.} All computed ESOP forms have been verified against their specifications, i.e., we simulated all ESOP forms for all possible values and compared the results of simulation with the initial truth tables of the provided Boolean functions.  Note that it is not possible to verify the minimality of the ESOP forms.

\noindent\textbf{NPN$4$.} We synthesized all ESOP forms of minmum size for all 222 representatives of the NPN$4$ equivalence classes~\cite{GotoT62}.  Computing one minimum ESOP form for each representatives takes $1.6$s, computing all minimum ESOP forms for each representatives takes $9.2$s.  Fig.~\ref{fig:npn4} shows the histogram of the size of the minimum ESOP forms for the representatives (on the left) and the number of ESOP forms of minimum size per representative (on the right).  On average a representative has 12 structurally different minimum ESOP forms.  Some representatives can have 100 or more ESOP forms of minimum size.  The Boolean function \texttt{0x166A} (shown in Fig.~\ref{fig:kvmap}) has the most minimum ESOP forms (in total $126$) within the NPN$4$ classes.

\noindent\textbf{LUT mapping.} We synthesized one ESOP form for a fixed number of ESOP terms and one ESOP form of minimum size using downward and upward search, respectively, for each Boolean function that occurred in LUT mapping of the EPFL benchmark suite.  For LUT mapping, we used the ABC command {\tt if -K 8}~\cite{BraytonM10}.  After LUT-mapping, we applied {\tt exactmine}~\cite{SoekenR18} to extract all Boolean functions from the benchmarks.  We obtained $4001$ different Boolean functions with up to $8$ Boolean variables and used SAT-based ESOP synthesis to compute ESOP forms.  For this experiment, we consider a fixed conflict limit of $10000$.  The synthesis results are presented in Table~\ref{tab:lut mapped}: the first column (\textbf{Terms}) is a user-specified upper limit on the number of terms.  The rest of the table is organized in three parts.  The first part (\textbf{fixed-size}) is dedicated to synthesis of an ESOP form for the given term limit (without minimizing the number of terms).  In this case, the SAT-solver's heuristics decides whether unecessary terms are cancelled or kept.  The second part (\textbf{downward search}) is dedicated to a synthesis procedure that iteratively synthesizes ESOP forms staring from the upper term limit and decreases the number of terms until the constraint system becomes unsatisfiable (as described in Algorithm~\ref{alg:exact cegar esop}).  The satisfying assignment with the smallest number of terms is used for deriving an ESOP form.  The last part (\textbf{upward search}) is similar to the second part, but starts with $1$ term and increases the number if the constraint system is unsatisfiable.  The satisfying assignment with the largest number of terms is used to derive an ESOP form.  For each part, we list the number of Boolean functions successfully realized (\textbf{R}), the number of Boolean functions that could not be synthesized because the SAT-solver's conflict limit (\textbf{C}) was exceeded, the average number of terms (\textbf{k}) for all realizable Boolean functions, and the total run-time (\textbf{T}) for synthesizing all Boolean functions.  The run-time includes the time for synthesizing the realizable Boolean function and the time spend in unsuccessful synthesis attempts.

\begin{table}[t]
  \centering
  \caption{Synthesis of ESOP forms for LUT mapping.}
  \label{tab:lut mapped}
  \def\tabcolsep{3.0pt}
  \begin{tabular}{c rrrr rrrr rrrr}
    \toprule
    \multicolumn{1}{c}{\bf Terms} & \multicolumn{4}{c}{\bf Fixed-size} & \multicolumn{4}{c}{\bf Downward search} & \multicolumn{4}{c}{\bf Upward search}\\
    \cmidrule(lr){2-5} \cmidrule(lr){6-9} \cmidrule(lr){10-13}
    &
    \multicolumn{1}{c}{\bf R} & \multicolumn{1}{c}{\bf C} & \multicolumn{1}{c}{\bf k} & \multicolumn{1}{c}{\bf T} &
    \multicolumn{1}{c}{\bf R} & \multicolumn{1}{c}{\bf C} & \multicolumn{1}{c}{\bf k} & \multicolumn{1}{c}{\bf T} &
    \multicolumn{1}{c}{\bf R} & \multicolumn{1}{c}{\bf C} & \multicolumn{1}{c}{\bf k} & \multicolumn{1}{c}{\bf T} \\
    \midrule
     8 & 3735 & 266 & 5.19 & 49.65s & 3854 & 147 & 3.60 &  300.44s & 3857 & 3854 & 3.60 & 248.07s \\
    16 & 3806 & 195 & 7.10 & 50.56s & 3965 &  36 & 3.82 &  695.08s & 3965 &   36 & 3.82 & 338.72s \\
    32 & 3966 &  35 & 8.45 & 42.67s & 4001 &   0 & 3.94 & 1430.41s & 4001 &    0 & 3.94 & 355.49s \\
    \bottomrule
  \end{tabular}
\end{table}

\begin{figure}[t]
  \begin{center}
    \pgfplotsset{every tick label/.append style={font=\tiny}}
    \pgfplotsset{scaled y ticks=false}
    \begin{tikzpicture}
      \begin{axis}[ymode=log, name=Histogram3, axis x line=bottom, xmin=0, xmax=17, ymin=0, ymax=600000, area style, xtick={0,1,2,3,4,5,6,7,8,9,10,11,12,13,14,15,16}, ytick={}, width=13cm, height=6cm, x tick label style={xshift=10pt}]
        \addplot+[ybar interval,forget plot,thick,blue,cyan!20!black,fill=cyan!80!black,mark=none] coordinates {(5,23928) (6,322792) (7,322792)};
        \addplot+[ybar interval,forget plot,thick,blue,cyan!20!black,fill=gray,mark=none] coordinates {(7,503805) (8,500435) (9,500435)}; % last element is sentinel
        \addplot+[ybar interval,forget plot,thick,green!20!black,fill=green!60!black,mark=none] coordinates {(10,45212) (11,45212)}; % last element is sentinel
        \addplot+[ybar interval,forget plot,thick,green!20!black,fill=green!60!black,mark=none] coordinates {(12,188393) (13,113363) (14,113363)}; % last element is sentinel
        \addplot+[ybar interval,forget plot,thick,blue,cyan!20!black,fill=cyan!80!white,mark=none] coordinates {(0,0) (1,0) (2,12) (3,163) (4,1853) (5,1853)}; % last element is sentinel
        \addplot+[ybar interval,forget plot,thick,green!20!black,fill=green!90!black,mark=none] coordinates {(9,61) (10,61)}; % last element is sentinel
        \addplot+[ybar interval,forget plot,thick,green!20!black,fill=green!90!black,mark=none] coordinates {(11,95) (12,95)}; % last element is sentinel
        \addplot+[ybar interval,forget plot,thick,green!20!black,fill=green!90!black,mark=none] coordinates {(14,117) (15,55) (16,134) (17,134)}; % last element is sentinel        
        \addplot+[mark=none, black, dashed] coordinates {(0,10000) (17,10000)};
      \end{axis}
    \end{tikzpicture}
  \end{center}
  \caption{SAT-solver results for different $k$ for {\tt 0xF550311031100000}.}
  \label{fig:sat-solver results}
\end{figure}
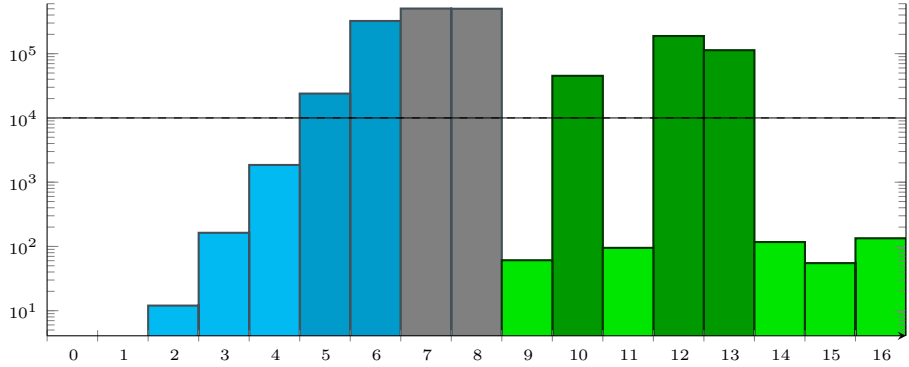

\begin{example}
We illustrate the effect of the conflict limit on upward and downward search with a simple example.  Consider the completely-specified Boolean function {\tt 0xF550311031100000}.  We attempt to synthesize an ESOP form of minimum size with at most $16$ terms and a conflict limit of $10000$ using the upward and downward search procedure, respectively.  Fig.~\ref{fig:sat-solver results} shows the number of conflicts explored by the SAT-solver in logarithmic scale parametrized by the number of terms (\textbf{k}).  The colors encode the decision results: green denotes satisfiable, blue denotes unsatisfiable, and gray denotes unknown.  For those $k$ for which the conflict limit of $10000$ was reached, we repeated synthesis with a much higher conflict limit of $500000$ to understand what conflict limit would allow us to conclude the correct result.  The results for $k=7$ and $k=8$, however, remain unknown, i.e., we do not know whether the constraints are satisfiable, because the conflict limit of $500000$ was also exceeded.

The downward search starts with $16$ terms and systematically decreases the number of terms.  During the search, the conflict limit is reached with $k=13$ for the first; the search procedure interprets this as potentially satisfiable, such that the procedure proceeds until finally $k=4$ is reached.  For $k=4$, the procedure concludes unsatisfiability, terminates, and returns the smallest constructed ESOP form with $9$ terms determined during the search process.

The upward search procedure solves the constraint system with increasing number of terms starting with $1$.  For $k \leq 4$, the SAT-solver proves unsatisfiability of the constraint system.  For $5 \leq k \leq 8$, the SAT-solver reaches the conflict limit, which is interpreted as potentially unsatisfiable by our search procedure, such that the search procedes until $k=9$.  For $k=9$ terms, the constraint system becomes for the first time satisfiable and the corresponding ESOP form with $9$ terms is returned.
\end{example}

\begin{table}[t]
  \centering
  \caption{Synthesis of ESOP forms for randomly-generated Boolean functions.}
  \label{tab:random}
  \def\tabcolsep{2.8pt}
  \begin{tabular}{cc rrrr rrrr rrrr}
    \toprule
    \multicolumn{1}{c}{\bf Var.} & \multicolumn{1}{c}{\bf Terms} &
    \multicolumn{4}{c}{\bf Fixed-size} & \multicolumn{4}{c}{\bf Downward search} & \multicolumn{4}{c}{\bf Upward search}\\
    \cmidrule(lr){3-6} \cmidrule(lr){7-10} \cmidrule(lr){11-14}
    & &
    \multicolumn{1}{c}{\bf R} & \multicolumn{1}{c}{\bf C} & \multicolumn{1}{c}{\bf k} & \multicolumn{1}{c}{\bf T} &
    \multicolumn{1}{c}{\bf R} & \multicolumn{1}{c}{\bf C} & \multicolumn{1}{c}{\bf k} & \multicolumn{1}{c}{\bf T} &
    \multicolumn{1}{c}{\bf R} & \multicolumn{1}{c}{\bf C} & \multicolumn{1}{c}{\bf k} & \multicolumn{1}{c}{\bf T} \\
    \midrule
     5 & 16 & 100 &  0 &  8.58 &  0.11s & 100 &  0 &  3.34 &    1.35s & 100 & 0 &   3.34 &    0.12s \\
     6 & 16 &  99 &  1 & 11.32 &  0.42s & 100 &  0 &  5.62 &   24.70s & 100 & 0 &   5.62 &   15.99s \\
     7 & 32 &  86 & 14 & 24.91 &  3.71s & 100 &  0 & 17.96 &  276.70s & 100 & 0 &  17.96 &  210.02s \\
     8 & 96 &  79 & 21 & 54.35 & 19.64s & 100 &  0 & 45.41 & 2156.96s & 100 & 0 &  45.41 & 1151.75s \\
     \bottomrule
  \end{tabular}
\end{table}

\noindent\textbf{Random.}
We synthesized ESOP forms for randomly-generated, in\-com\-plet\-ely-specified Boolean functions over $5$, $6$, $7$, and $8$ Boolean variables.  Each bit in the Boolean function and its care function was chosen by flipping a fair coin.  In total, we generated $100$ Boolean functions for each number of Boolean variables.  Table~\ref{tab:random} summarizes the results for synthesizing ESOP forms.  The first two columns list the number of Boolean variables (\textbf{Var.}) and a fixed bound on the number of terms (\textbf{Terms}).  The rest of the table is organized as Table~\ref{tab:lut mapped}.  Due to the symmetric design of downward and upward search, they reached exactly the same minimum ESOP forms.  Overall downward search is slower due to the fact that unsatisfiability is typically harder to prove and can only be concluded by the SAT-solver for sufficiently small $k$.   Consequently, the downward search procedure on average analyzes many more cases before unsatisfiability is reached.  In contrast, upward search keeps searching until satisfiability is reached for the first time, which can occur early in the search process.

\section{Conclusion}

We have presented an exact synthesis approach for computing ESOP forms using Boolean satisfiability.  The approach needs no pre-computed information, synthesizes one or multiple ESOP forms of minimum size, and can take completely-specified or in\-com\-ple\-tely-spe\-ci\-fied Boolean functions as specifications.  We have implemented the approach using an off-the-shelf SAT-solver and have further presented an relaxation that leverages the SAT-solver's conflict limit to find ESOP forms with almost minimum size.  We have also presented evidence that the synthesis procedure can deal with small-scale ESOP forms with up to $8$ Boolean variables and up to $100$ terms.  As benchmarks, we have used Boolean functions in the NPN$4$ equivalence class, Boolean functions that appeared during $8$-LUT mapping, and randomly-generated Boolean functions.  We envision that the proposed SAT-based synthesis technique can be integrated with large-scale ESOP optimization procedures, e.g., by selecting windows of terms and resynthesizing them.

\section{Acknowledgements}
This research was supported by H2020-ERC-2014-ADG 669354 CyberCare (200021-146600) and the Institutional Strategy of the University of Bremen, funded by the German Excellence Initiative.

\end{document}